\documentclass[prb,twocolumn,superscriptaddress,showpacs]{revtex4}
\usepackage{amssymb}
\usepackage{txfonts}
\usepackage{bbm}
\usepackage{graphicx,graphics,color,epsfig}
\usepackage{appendix}
\usepackage{epstopdf}
\usepackage{appendix}

\begin{document}

\title{Anomalous sharp peak in the London penetration depth induced by the nodeless-to-nodal superconducting transition in
BaFe$_2$(As$_{1-x}$P$_x$)$_2$}

\author{Huai-Xiang Huang}
\email{hxhuang@shu.edu.cn}
\affiliation{Department of Physics, Shanghai University, Shanghai 200444,
China}

\author{Wei Li}
\email{w_li@fudan.edu.cn}
\affiliation{State Key Laboratory of Surface Physics and Department of Physics, Fudan University, Shanghai 200433, China}
\affiliation{Collaborative Innovation Center of Advanced Microstructures, Nanjing University, Jiangsu 210093, China}

\author{Yi Gao}
\affiliation{Department of Physics and Institute of Theoretical Physics, Nanjing Normal University, Nanjing, Jiangsu 210023, China}

\author{Yan Chen}
\affiliation{State Key Laboratory of Surface Physics and Department of Physics, Fudan University, Shanghai 200433, China}
\affiliation{Collaborative Innovation Center of Advanced Microstructures, Nanjing University, Jiangsu 210093, China}

\author{Fu-Chun Zhang}
\affiliation{Kavli Institute for Theoretical Sciences and CAS Center for Excellence in Topological Quantum Computation, University of Chinese
Academy of Sciences, Beijing 100190, China}
\affiliation{Collaborative Innovation Center of Advanced Microstructures, Nanjing University, Jiangsu 210093, China}

\date{\today}
\begin{abstract}
The issue of whether the quantum critical point (QCP) is hidden inside unconventional superconductors is a matter of hot debate. Although a
prominent experiment on London penetration depth has demonstrated the existence of the QCP in the isovalent-doped iron-based superconductor
BaFe$_2$(As$_{1-x}$P$_x$)$_2$, with the observation of a sharp peak in the penetration depth in the vicinity of the disappearance of magnetic order
at zero temperature, the nature of such an emerging QCP remains unclear. Here, we provide a unique picture to understand well the phenomena of the
QCP based on the framework of linear response theory. Evidence from the density of states and superfluid density calculations suggests the
nodeless-to-nodal pairing transition accompanied the appearance of a sharp peak in the penetration depth in BaFe$_2$(As$_{1-x}$P$_x$)$_2$. Such a
pairing transition originates from the three-dimensional electronic properties with a strong interlayer superconducting pairing. This finding
provides a significant insight into the understanding of the QCP observed in experiment in BaFe$_2$(As$_{1-x}$P$_x$)$_2$.
\end{abstract}

\maketitle

\section{Introduction}
Studies of unconventional iron-based superconductivity have triggered intensive research interests during the past decade since the
discovery of LaO$_{1-x}$F$_x$FeAs in 2008~\cite{kam}. For low-energy electronic properties, iron-based materials are a multiband system
with nodeless $s_{\pm}$-wave superconducting pairing symmetry~\cite{HDing,GRStewart} in contrast to that of cuprates, which are a single band
system with nodal $d$-wave superconducting pairing symmetry~\cite{Damascelli,CCTsuei}. Despite such differences at the microscopic level, the
layered crystal structure and phase diagram of both iron-based and copper-oxide superconductors share a common feature. From the viewpoint of
the superconducting phase diagram, those compounds exhibit similar dome-shaped superconductivity after introducing the extra electron or
hole-like charge carriers into the parent compound or applying high external pressure/or chemical pressure. An isovalent phosphorus
substitution of arsenic in the BaFe$_2$(As$_{1-x}$P$_x$)$_2$ compound accompanied by the appearance of
superconductivity~\cite{Rotter,Shishido,Kasahara,yut,hsu,mdz} can be regarded as a kind of chemical pressure. Importantly, a prominent
experiment on London penetration depth in this compound observed a sharp peak in the vicinity of the disappearance of magnetic order at zero
temperature, suggesting the presence of quantum critical point (QCP)~\cite{Hashimoto} and attracting widespread research
attention~\cite{yna,rafa,zdiao,Smylie,add1}.

Elucidating the origin of such QCP inside superconducting dome could be the key to understanding high temperature
superconductivity~\cite{add1,dai,mat,ale,yla,dze,add2}. Since the parent compound of BaFe$_2$As$_{2}$ has a collinear antiferromagnetic order, tuning the
electronic band structure by introducing isovalent phosphorus dopants without introducing charge carriers will suppress the magnetic order and
superconductivity will emerge. This leads to conjecture regarding whether the disappearance of magnetic order will be associated with a sharp peak in the
superfluid density in the London penetration depth experiment~\cite{Hashimoto,yna}. A previous theoretical study demonstrated that in two
dimensional systems the concentration of superfluid density, which is proportional to London penetration depth, $\rho_s\propto 1/\lambda^2_L$,
monotonically increases with the suppression of the magnetic order in the region where magnetism and superconductivity coexist, until the superfluid
density saturates to a maximal value in a pure superconducting region in Fe-based superconductors~\cite{huang2}. Therefore, such conjecture
seems to be insufficient to explain the nature of the London penetration depth experiment, and various theoretical scenarios are proposed
to explain the possible nature of such an anomalous enhancement of $\lambda_L$~\cite{add1,chow,nom,wal}.

Fortunately, angle-resolved photoemission spectroscopy (ARPES) measurements on the superconducting gap structure of
BaFe$_2$(As$_{0.7}$P$_{0.3}$)$_2$ demonstrated the direct observation of a circular line node on the most significant hole Fermi surface around
the $Z$ point at the Brillouin zone boundary~\cite{YZhang}. This finding opens an avenue for conjecturing whether the QCP observed in the
penetration depth experiment is closely related to such nodal pairing structure. In addition, the ARPES experiment and the first-principles
calculations also suggested that the Fermi surface topology becomes much more three-dimensional with increasing the phosphorus
dopants~\cite{Shishido,YZhang,Hashimoto,liwei,KSuzuki2011}, leading us to establish a perspective of the nodeless-to-nodal pairing transition
accompanied by the appearance of QCP in BaFe$_2$(As$_{1-x}$P$_x$)$_2$, which is the primary motivation of the present paper.

In this paper, a doping-dependent three-dimensional tight-binding model is constructed to reproduce well the correct low-energy electronic
band structure and the Fermi surface topologies from ARPES measurements~\cite{Shishido}. By taking the Coulomb interactions between itinerant
electrons into account, we perform self-consistent mean-field calculations and obtain a phase diagram of pairing order parameters versus
doping concentrations, which is in agreement with experiments~\cite{Hashimoto}. Further calculations of superfluid density and the density of
states (DOS) as a function of doping demonstrate that the appearance of a sharp peak in the penetration depth is accompanied by a nodeless-to-nodal
pairing transition. Such a superconducting pairing transition mainly comes from the nature of the three-dimensional electronic band structure with
strong interlayer superconducting pairing order.  Additionally, it is worthy pointing out that the calculated maximum $\lambda_L$ does not appear at the transit point of magnetic order observed by experiment~\cite{Hashimoto}, instead it is within the overlapped range of spin-density-wave and superconducting phases. The same feature was reported in a previous work [\onlinecite{add1}] by using the universal critical phenomena theory, which indicated that the possible explanation of the discrepancy between experiment observation and theoretical calculation requires the consideration of the physical properties at the scale of the correlation length or an even smaller length scale.

The rest of this paper is organized as follows. In Sec.~\ref{Sec2}, we first introduce the theoretical model Hamiltonian and the methods of the
detailed calculations. The calculated phase diagram, superfluid density, and London penetration depth at zero temperature are given in
Sec.~\ref{results}. In Sec.~\ref{dos}, the DOS and Fermi surface of the superconducting state are addressed. A summary is finally given in
Sec.~\ref{final}.

\section{ Model Hamiltonian}\label{Sec2}

\begin{figure}
\centering
     \centering
      \includegraphics[width=4.5cm]{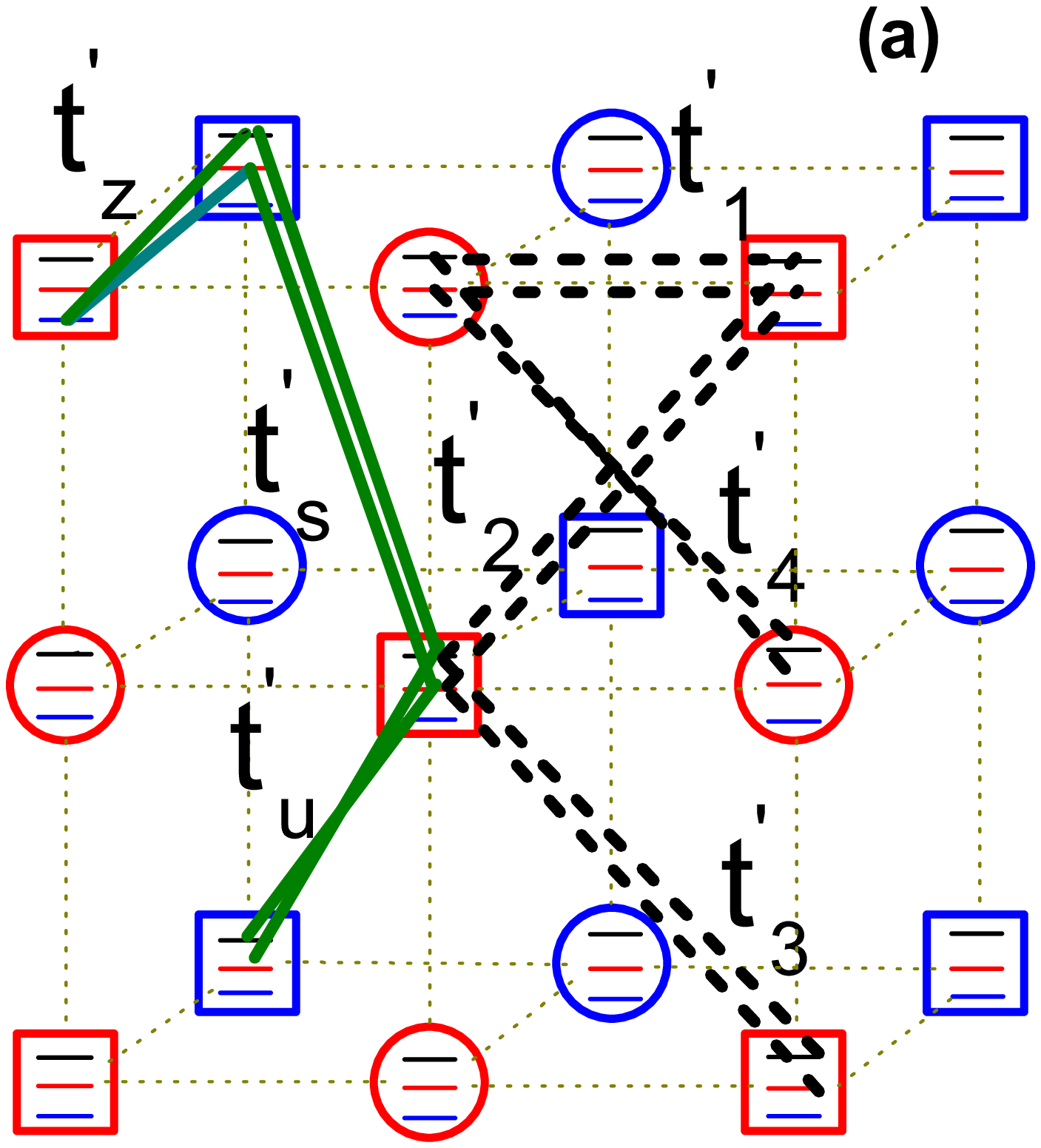}
      \includegraphics[width=3.5cm]{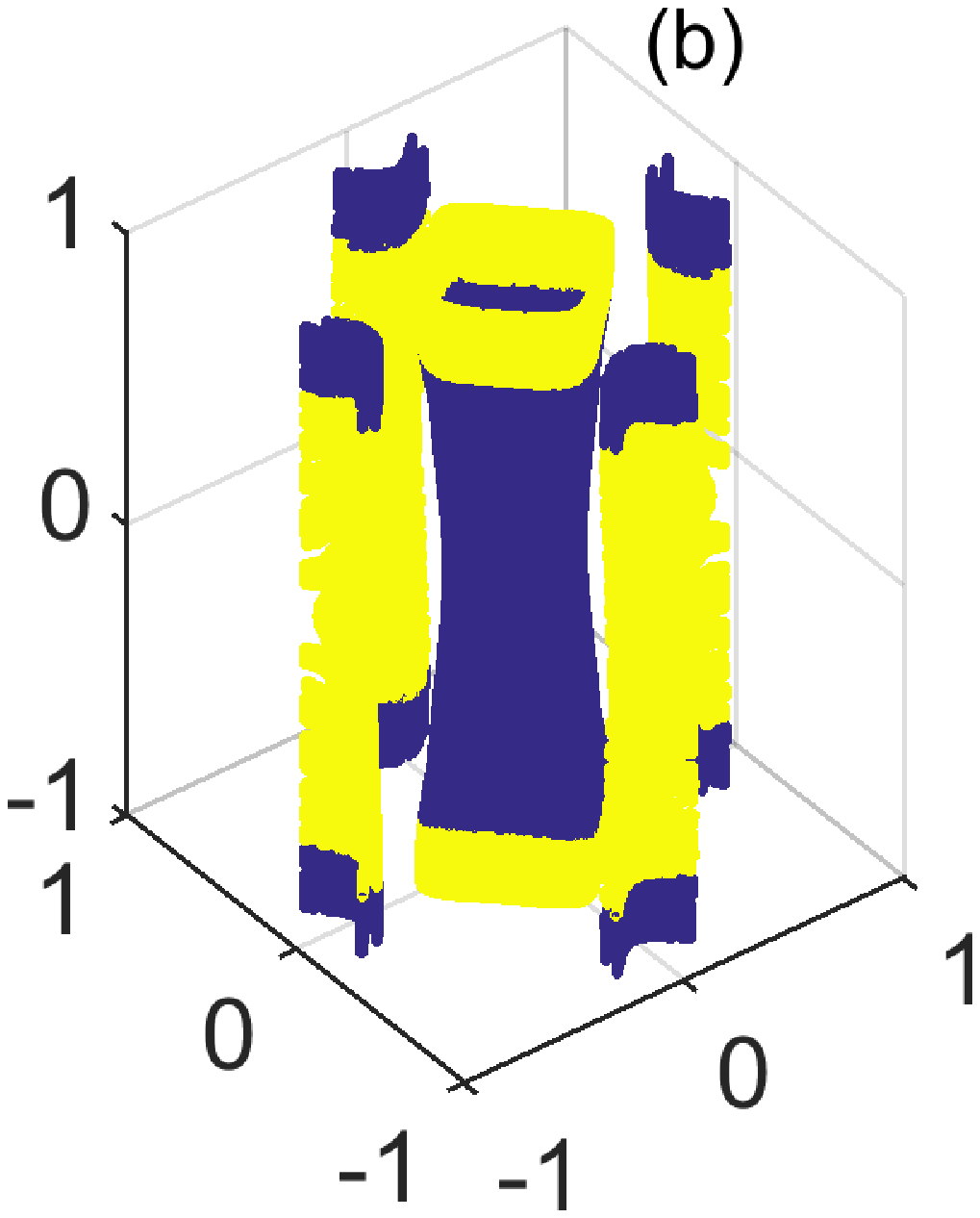}
      \includegraphics[width=4.0cm]{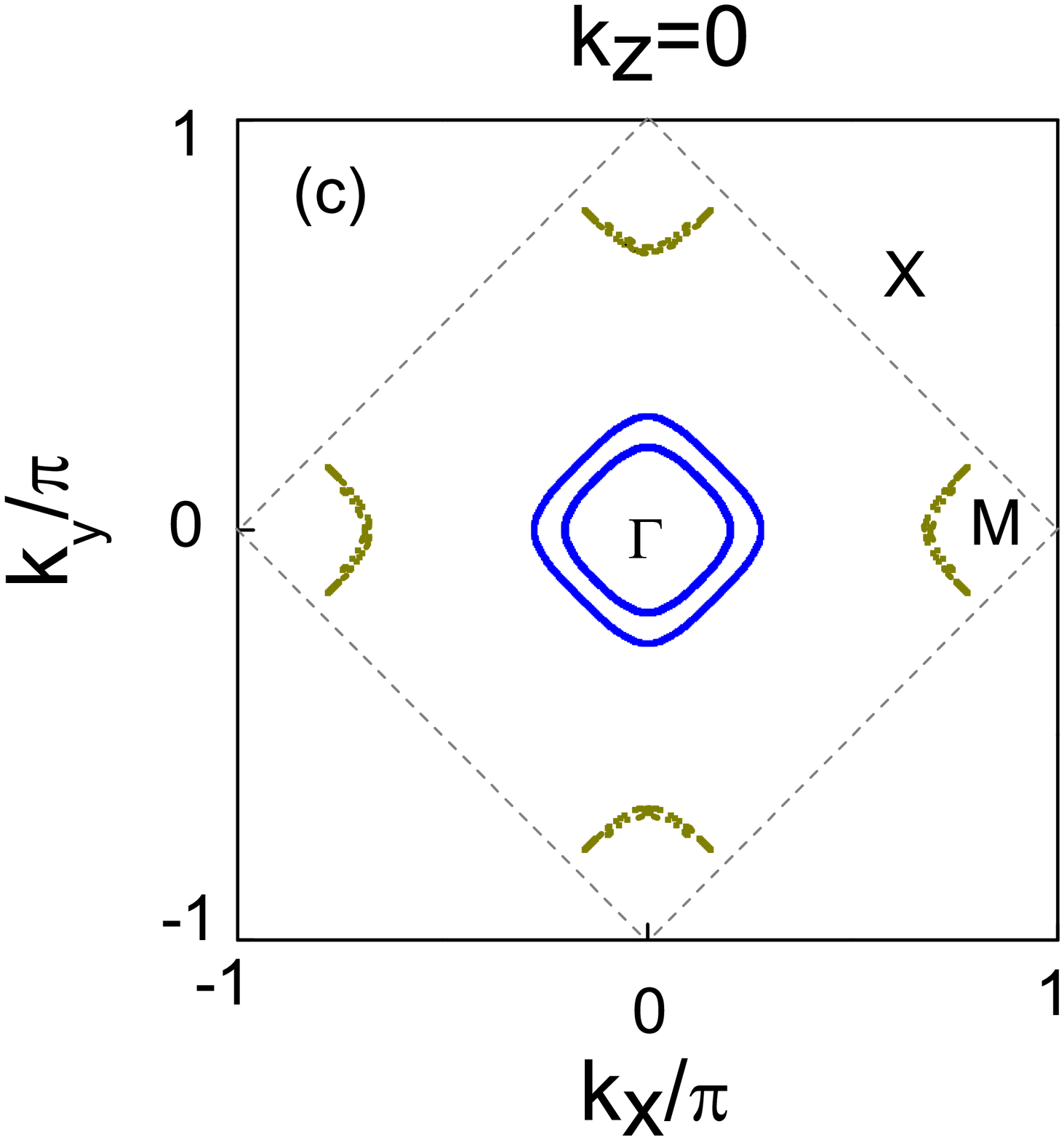}
      \includegraphics[width=4.0cm]{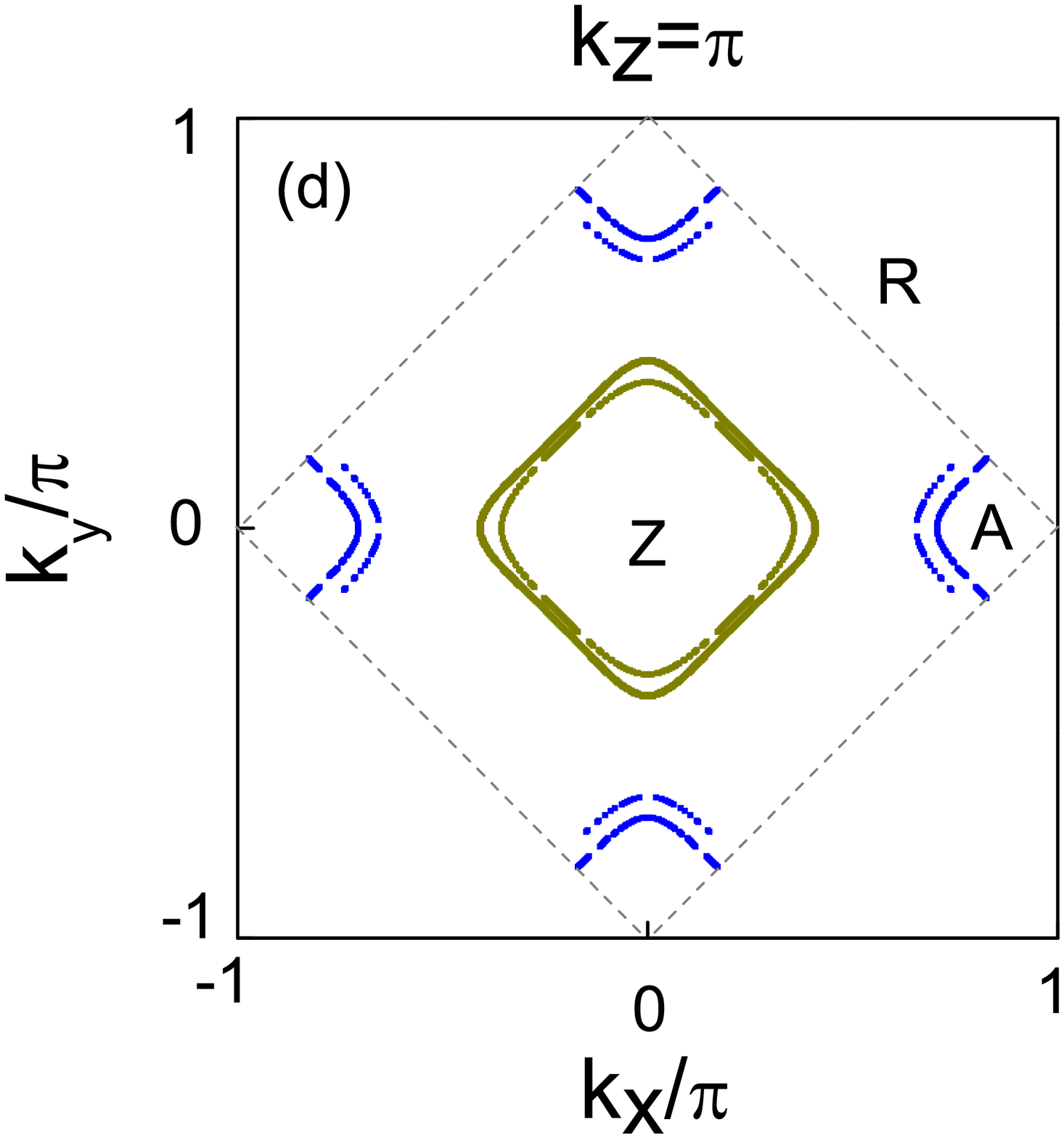}
      \includegraphics[width=4.0cm]{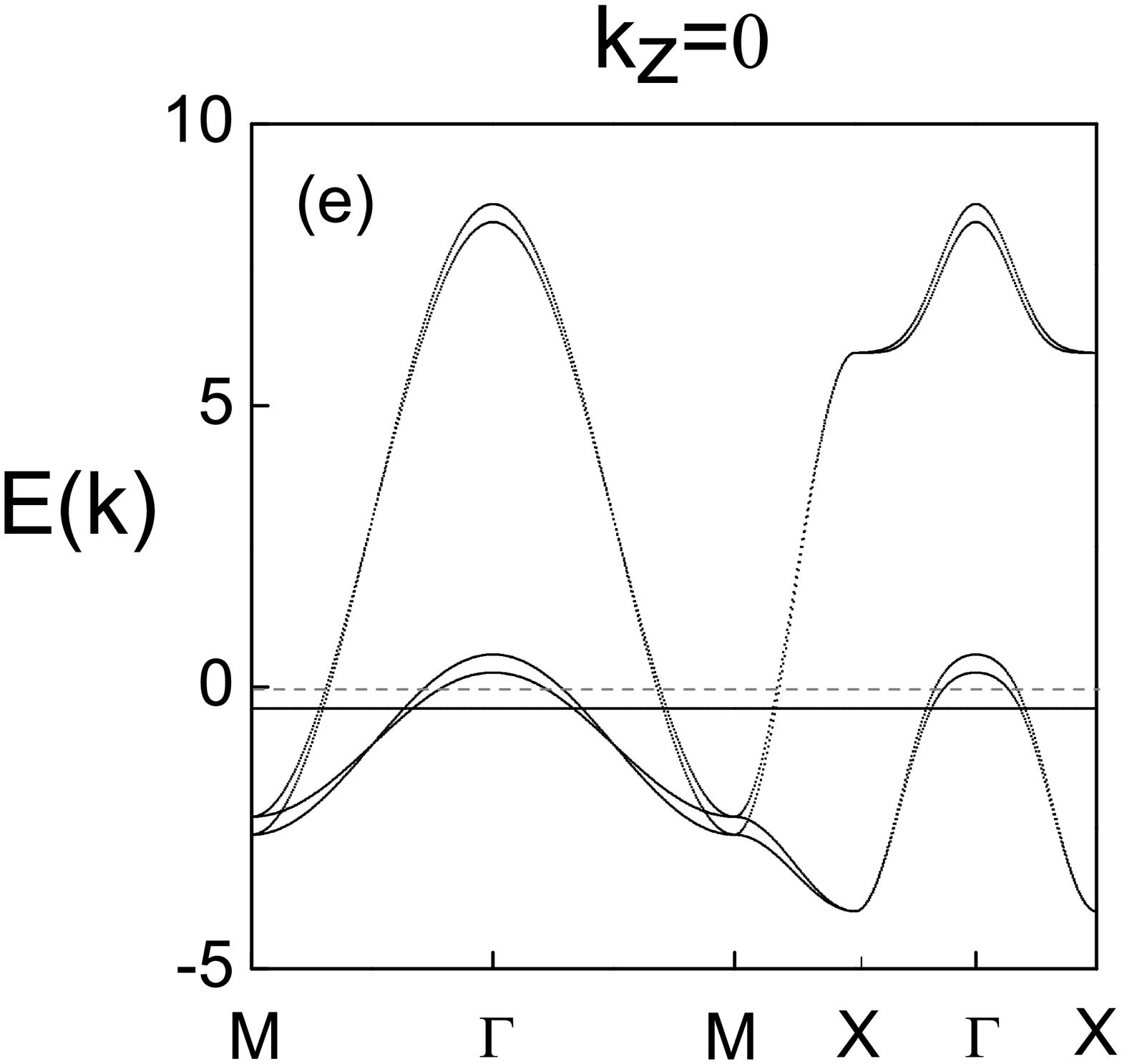}
      \includegraphics[width=4.0cm]{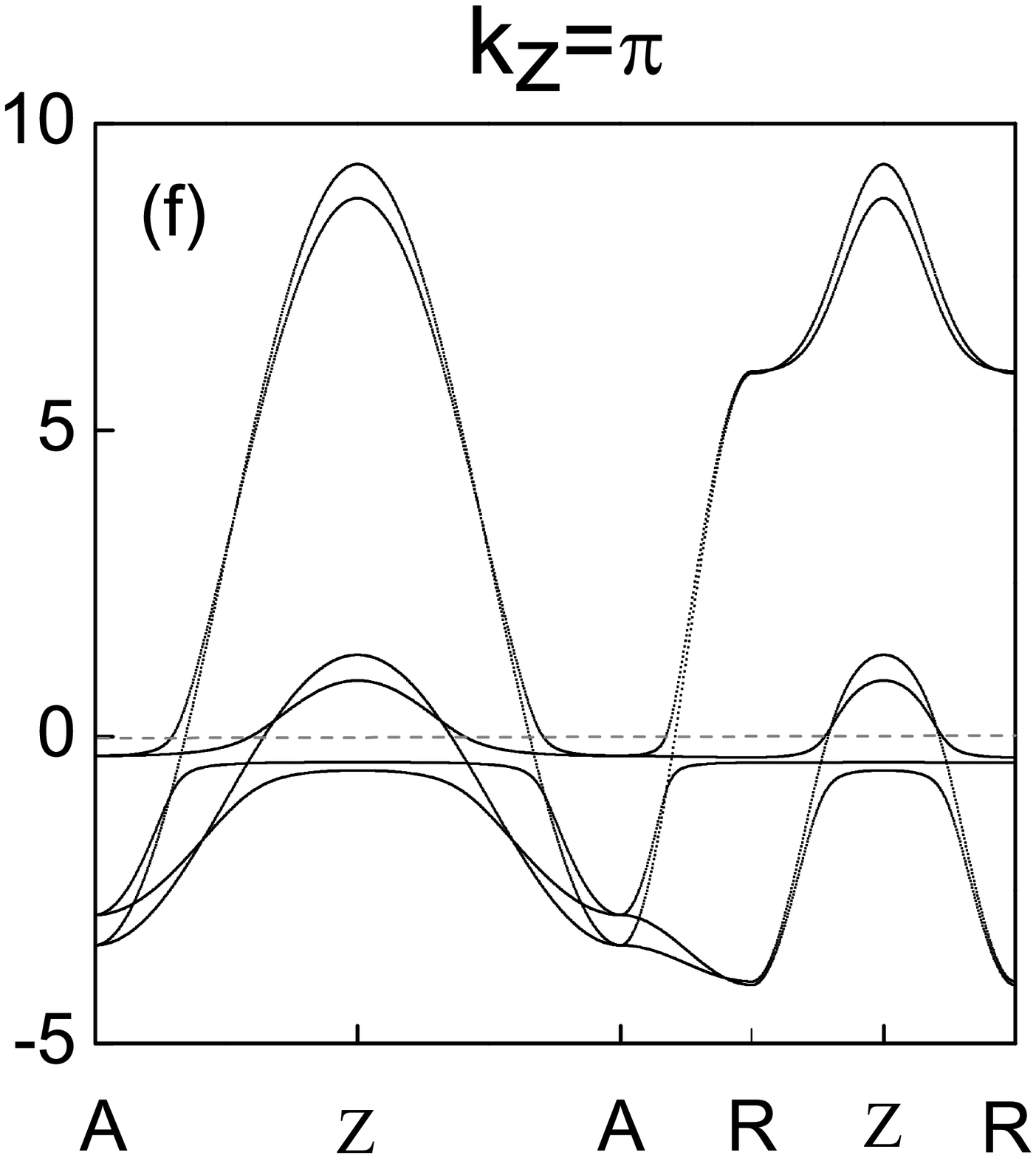}
\caption{(Color online) (a) Schematic illustration of hopping energy parameters of the three dimensional tight-bind model
$-t_{\mathrm{ij}}c^{\dag}_{\mathrm{i}}c_{\mathrm{j}}$ [$\mathrm{i}$ denotes all the index of the site $i$] with three Fe orbitals $d_{xz}$,
$d_{yz}$ and $d_{z^2}$ in real space. The red symbols denote the sites in one Fe layer, and the blue symbols denote those in the adjacent Fe layer.
The two inequivalent Fe atoms are denoted by circles and squares. Hopping in the same layer is linked by black dashed lines, while
hopping between adjacent layers is linked by solid olive ones. (b) The three-dimensional Fermi surface topologies of
BaFe$_2$(As$_{1-x}$P$_x$)$_2$ for $x=0.18$. The label of the space axes is $k_{i=x,y,z}/\pi$. For the three dimensional Fermi surface
topologies, blue (yellow) points represent plus (minus) signs of superconducting order in the band space, which appears after introducing the
interactions. The contour plot of the three-dimensional Fermi surface structure (c) for $k_z=0$ and (d) for $k_z=\pi$. The dashed lines
denote the first Brillouin zone with two Fe atoms per unit cell. The band structure is plotted along the high symmetric $\mathbf{k}$-points
for (e) $k_z=0$ and (f) $k_z=\pi$. The gray dashed line denotes the Fermi level.}\label{fig1}
\end{figure}

According to the fact of experimental measurements~\cite{Shishido,Hashimoto}, we extend the two-dimensional phenomenological model with
two orbitals~\cite{zhang}
to a three-dimensional model with three orbitals to study the superconducting electronic properties in  isovalent-doped
BaFe$_2$(As$_{1-x}$P$_x$)$_2$. The previous two-dimensional model considered the effect of asymmetric arsenic atoms is appropriate to
describe the experimental observations in ARPES and scanning tunnel microscope for the 122 family~\cite{huang4,huang3,jian1,yi4}.
The calculated superfluid density is in qualitative agreement with the direct experimental measurement in films of Fe pnictide superconductors
at low temperatures~\cite{Jyong}. In the extended model, a unit cell contains two Fe atoms, and each Fe involves three orbitals $d_{xz}$,
$d_{yz}$ and $d_{z^2}$. As arsenic is gradually substituted by phosphorus, the $d_{z^2}$ orbital of Fe will be driven close to the Fermi
level~\cite{Yueh,liwei}, resulting in an enhancement of interlayer hybridization between two interlayer Fe orbitals.

In Fig.~\ref{fig1}(a), we show a schematic illustration of tight-binding model Hamiltonian in real space, where $t^{\prime}_{1-4}$ are hopping
energies within each Fe layer between $d_{xz}$ and $d_{yz}$ orbitals. Here, it should be noted that $t^{\prime}_2$ is different from $t^{\prime}_3$ since asymmetric arsenic ions is above and below the Fe layer alternatively~\cite{zhang}. $t^{\prime}_{z,s,u}$ are the
interlayer hopping energies between two adjacent Fe layers. $t^{\prime}_z$ denotes the nearest-neighbor hopping energy along the $z$ axis between
$d_{z^2}$ and $d_{xz}$($d_{yz}$) orbitals, which can be regarded as two-step hopping processes
$c^{\dag}_{i+z}A_{i+z/2}A^{\dag}_{i+z/2}c_{i}+H.c.$ mediated by arsenic (denoted by A) in the crystal structure environment of
BaFe$_2$(As$_{1-x}$P$_x$)$_2$.
Under the $C_2$ rotation at site $i$, $c^{\dag}_{i+z,\beta\sigma}A_{i+z/2}\longrightarrow -c^{\dag}_{i-z,\beta}A_{i-z/2\sigma}$, the
combination of $c^{\dag}_{i+z,\beta\sigma}A_{i+z/2}-c^{\dag}_{i-z,\beta\sigma}A_{i-z/2}$ will replace the hopping term
$c^{\dag}_{i+z}A_{i+z/2}$~\cite{Yueh,wen}, and thus the two step hopping processes become $-4t_z\sin^2{k_z}$ after omitting the creation and
annihilation operators of arsenic.
$t^{\prime}_s$($t^{\prime}_u$) is the hopping energy along $\hat{x}\pm\hat{y}\pm \hat{z}$ between the same (different) $d_{xz},
d_{yz}$ orbitals in two adjacent Fe layers. Using the Fourier transformation, the tight-binding model Hamiltonian in momentum space can be rewritten as:
\begin{eqnarray}\label{htk}
H_{t,k}\!\!&=&\!\!\sum_{k\nu\sigma R} a_1 c^{\dag}_{A\nu\sigma k}c_{A\nu\sigma k}+a_2 c^{\dag}_{B\nu\sigma k}c_{B\nu\sigma
k}+a_3c^{\dag}_{R\nu\sigma k}c_{R\bar{\nu}\sigma k}\nonumber\\
&+&a_4c^{\dag}_{A\nu\sigma k}c_{B\nu\sigma k}+a_5 c^{\dag}_{R\nu\sigma k}c_{R\beta\sigma k}+H.c.,
\end{eqnarray}
where $R=A, B$ denotes two inequivalent sites of Fe atoms, $\nu$ denotes the orbitals of $d_{xz}, d_{yz}$, $\beta$ is the $d_{z^2}$ orbital,
and $\sigma$ is the spin. Comparing hopping terms in k space with the hopping parameters in real space, we obtain the coefficients in model Hamiltonian (\ref{htk}) as
follows
\begin{eqnarray}
a_1&=&-2t_2\cos{(k_x+k_y)}-2t_3\cos{(k_x-k_y)}\nonumber\\
&&-2t_s(1-\cos{k_z})(\cos{(k_x+k_y)}+\cos{(k_x-k_y)}),\nonumber\\
a_2&=&-2t_3\cos{(k_x-k_y)}-2t_2\cos{(k_x+k_y})\nonumber\\
&&-2t_s(1-\cos{k_z})(\cos{(k_x+k_y)}+\cos{(k_x-k_y)}),\nonumber\\
a_3&=&-2t_4(\cos{(k_x+k_y)}+\cos{(k_x-k_y)})\nonumber\\
&&-2t_u(1-\cos{k_z})(\cos{(k_x+k_y)}+\cos{(k_x-k_y)}),\nonumber\\
a_4&=&-2t_1(\cos{k_x}+\cos{k_y}),\nonumber\\
a_5&=&-2t_z(1-\cos{k_z}),\nonumber
\end{eqnarray}
with $t_{2,3}=t^{\prime}_{2,3}-t_s, t_4=t^{\prime}_4-t_u$, $t_{z,s,u}=-2t^{\prime}_{z,s,u}$, $t_{1}=t^{\prime}_{1}$ as shown in
Fig.~\ref{fig1}(a).

Diagonalizing the model Hamiltonian (\ref{htk}), we plot the three-dimensional Fermi surface topology as shown in Fig.~\ref{fig1}(b). There are
two quasi cylindrical shells around the $\Gamma$ point and two quasi cylindrical shells around the $M$ point. Figure ~\ref{fig1}(b) also shows the variation
of the three
dimensional Fermi surface along the $z$ direction, which is quite different from that in LaOFeAs superconductors~\cite{IIMazin}. For more
detail, we depict the contour plots of the three dimensional Fermi surface for $k_z=0$ and $k_z=\pi$ in Figs.~\ref{fig1}(c) and (d),
respectively.
The two Fermi surface circles around the $\Gamma$ point are enlarged; in particular, the inner circle  grows significantly with increasing $k_z$
along
the $z$ direction, while the variation of cylindrical shells around the $M$ points is insignificant. Those low-energy electronic behaviors are
in
good agreement with previous ARPES measurements~\cite{Shishido,Hashimoto}. In addition, the corresponding electronic band structures
$E(k_x,k_y,k_z)$ are
plotted for $k_z=0$ and $k_z=\pi$ respectively, in Figs.~\ref{fig1}(e) and (f) along the high-symmetry $\mathbf{k}$-points.
For the tight-binding model, there are six bands, where the two bands of $d_{z^2}$ orbital are degenerate and dispersive below the Fermi level
for $k_z=0$ shown in Fig.~\ref{fig1}(e). However, for a finite $k_z$, the two degenerate flat bands will be split and become much more
dispersive, which can be seen clearly in Fig.~\ref{fig1}(f).

Taking the strong Coulomb interactions between itinerant electrons in Fe three-dimensional ($3D$) orbitals into account, we write the
interaction Hamiltonian on a mean-field level as $H_{int}=H^{xy}_{int}+H^{z}_{int}$, which is expressed as~\cite{zhou,ygao,amo}:
\begin{eqnarray}\label{2}
H^{xy}_{int}&=&U\!\!\sum_{i\sigma\nu}\!\langle n_{{
i}\nu\bar{\sigma}}\rangle n_{{i}\nu\sigma}+(U-3J_H)\sum_{
i\nu\sigma} \langle n_{i\nu\sigma}\rangle n_{i\nu\sigma}\nonumber\\
&&+(U-2J_H)\sum_{i\nu\sigma} \langle n_{
i\nu\bar\sigma}\rangle n_{i\nu \sigma}-\sum_{i\nu\sigma}\mu n_{i\nu \sigma},\\
\hat{H}^{z}_{int}&=&\sum_{i\sigma}U_z\langle n_{i\beta\sigma} \rangle n_{i\beta\bar{\sigma}} -\mu_1 n_{i\beta \sigma},
\end{eqnarray}
where the parameter $J_{H}$ denotes the Hund's coupling, and $U$ and $ U_z$ describe on-site Coulomb interaction on the $d_{xz}$($d_{yz}$) and
$d_{z^2}$ orbitals, respectively. Since the $d_{z^2}$ orbital is far below the Fermi level~\cite{liwei} in the parent compound BaFe$_2$As$_2$,
without loss of generality, we set $\mu_1=\mu+1.36$ and search $\mu$ self-consistently to fix the total electron number as a constant (4
electron/per Fe atom) throughout all calculations. All interactions and hopping parameters, such as $t_2$, $t_3$, and $t_z$, are doping
dependent with fixed relations of $t_s=-t_z$, $t_u=0.2t_z$, and $t_4=0.04$. The wave vector $\bf{k}$ is restricted in the magnetic Brillouin
zone, ascribed to the system displaying a spin-density wave order. In addition, the local electron density is expressed as
$n_{i\nu\sigma}=\frac{1}{4}\langle n_i\rangle+\sigma M_i$, and the magnetic order is described as
$M=\frac{1}{2}\sum_{\nu}(n_{A\nu\uparrow}-n_{A\nu\downarrow})=\frac{1}{2N_s}\sum_{\nu,\bf{k}} \sigma c^{\dag}_{A\nu\sigma
\bf{k}}c_{A\nu\sigma \bf{\bf{k+Q}}}$. Here $N_s$ is the number of unit cells, and $Q=(0,\pm\pi)$ or $(\pm\pi,0)$ is the wave vector of
spin-density wave order~\cite{Rotter}.

Furthermore, we consider the intralayer and interlayer superconducting pairings between the same $d_{xz}$ and $ d_{yz}$ orbitals as
$H_{\Delta}=\sum_{R\nu\tau\tau^{\prime}}\Delta^{R\nu}_{i,i+\tau}c^{\dag}_{i\nu
\uparrow}c^{\dag}_{i+\tau,\nu \downarrow}+h.c.$,
where $\tau=x\pm y$ and $\tau^{\prime}=x\pm y \pm z$.
In momentum space, the superconducting Hamiltonian reads $H_{\Delta,k}=\sum_{R\nu
\bf{k}}(\Delta_{R\nu\bf{k}}c^{\dag}_{R\nu
\bf{k}\uparrow}c^{\dag}_{R\nu -\bf{k}\downarrow}+h.c.)$, with
\begin{eqnarray}\label{dtk}
\Delta_{R\nu\bf{k}}=2\sum_{\tau}\cos{\bf{k}}_{\tau}\Delta^{R\nu}_{i,i+\tau}+2\sum_{\tau^{\prime}}\cos{\bf{k}}_{\tau^{\prime}}\Delta^{R\nu}_{i,i+\tau^{\prime}}\nonumber\\
=4\cos{k_x}\cos{k_y}(\Delta^s_{xy}+2\Delta_z\cos{k_z})-4\Delta^{d}_{xy}\sin{k_x}\sin{k_y},
\end{eqnarray}
where the self-consistent pairing order parameter $\Delta^{R\nu}_{i,i+\tau}=\frac{V_{\tau}}{2}\langle
c^{R}_{i\nu\uparrow}c^{R}_{i+\tau,\nu
\downarrow}-c^{R}_{i\nu\downarrow}c^{R}_{i+\tau,\nu
\uparrow} \rangle$ can be solved numerically. Interestingly, the value of superconducting pairing order within a Fe layer can be expressed as a
linear combination of $d$-wave and $s$-wave pairing orders defined by
$\Delta^{s,d}_{xy}=0.5(\Delta^{R\nu}_{i,i+\hat{x}+\hat{y}}\pm\Delta^{R\nu}_{i,i+\hat{x}-\hat{y}})$, because the superconducting pairing order
on $\hat{x}+\hat{y}$-orientated links is different from that on $\hat{x}-\hat{y}$-orientated ones. The interlayer pairing order is denoted $\Delta_z$
hereafter for short, the paring potential $V_{xy,z}=1.6$ for both intra- and interlayer superconducting pairing order. Here, it should be
noted that when the pairing order parameter $\Delta_z$ approaches zero, the three-dimensional superconductivity will evolve into an
exact two-dimensional superconducting system, and the pairing order $\Delta_{R\nu\bf{k}}$ has $s_{\pm}$ symmetry with the nodal lines located
at around $k_x=\pm\pi/2$ and $k_y=\pm\pi/2$. When the pairing order parameter $\Delta_z$ is increased to a finite value, such as
$|\frac{-\Delta^s_{xy}}{2\Delta_z}|\leq 1$, some extra nodal points will penetrate into the hole pockets at around the $\Gamma$ point.

In the numerical calculation, we set the distance between the nearest-neighbor Fe atoms and the hopping integral $t_1$ as the length and energy units,
respectively. By self-consistently diagonalizing the $24\times24$ total Hamiltonian in momentum space,
$H^0_{tot}(k)=H_{t,k}+H_{\Delta,k}+H_{int,k}=\sum_n E_n\gamma^{\dag}_n\gamma_n$, we obtain the eigenvalues and the corresponding eigenstates of
the system, which can be used for further calculating the physical quantities, such as superfluid density and the local DOS. The unit
cell is $128\times128\times128$ for the self-consistent calculation and $384\times384\times384$ for the calculations of superfluid density and
band structure, as well as DOS.

\section{Phase diagram and London penetration depth}\label{results}

\begin{figure}
\centering
      \includegraphics[width=9.0cm]{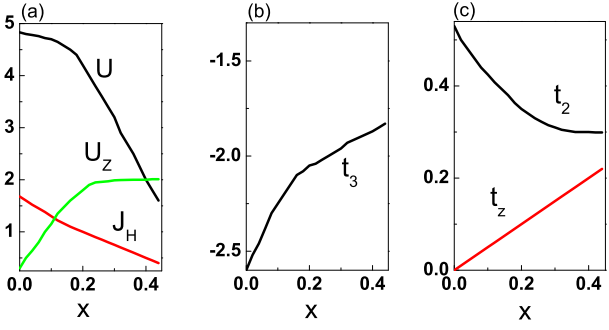}
\caption{(color online) Doping dependent (a) Coulomb interactions including U, $J_H$ and $U_z$, hopping energy parameters (b) $t_3$, and (c) $t_2$ and $t_z$. }\label{fig2}
\end{figure}

\begin{figure}
\centering
\includegraphics[width=9cm]{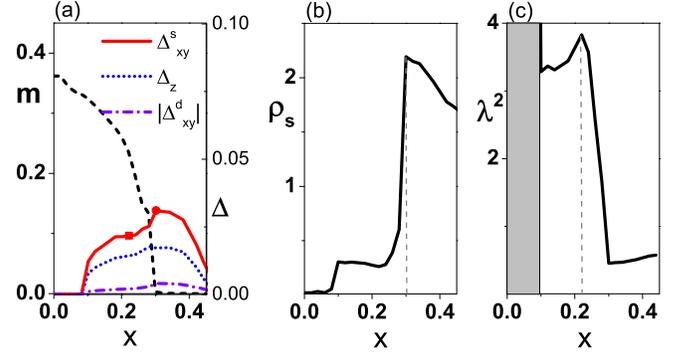}
\caption{(Color online) (a) Doping-dependent magnetic order $m$, intralayer superconducting orders $\Delta^{s,d}_{xy}$ and interlayer
superconducting order $\Delta_{z}$ for BaFe$_2$(As$_{1-x}$P$_x$)$_2$. (b) Superfluid density $\rho_s$ versus doping level $x$. (c) Square of the
penetration depth $\lambda^2$ versus $x$.
}\label{fig3}
\end{figure}

Figure~\ref{fig2} shows the doping dependent parameters used for the detailed calculations, including all hopping energies and interactions,
$t_2,t_3,t_z,U,U_z,J_H$ smoothly varied under various doping concentrations. The variation of these parameters is constructed only to fit the experimental
observations~\cite{YZhang,Shishido,Hashimoto}.
The numerically self-consistently calculated phase diagram is shown in Fig.~\ref{fig3}(a), which is in quit good agreement with previous
experimental observations~\cite{Hashimoto,yna,ding}. The parent compound BaFe$_2$As$_2$ with  antiferromagnetic order will be suppressed
monotonously with increasing the doping concentration $x$. When the doping is increased beyond $x=0.08$, the superconductivity emerges, as
evidenced by the appearance of the intralayer and interlayer superconducting pairing order parameters $\Delta^{s,d}_{xy}$ and $\Delta_{z}$,
and then the system enters the region where magnetism and superconductivity coexist until the doping concentrations reaching $x=0.3$. If we
further increase the doping concentrations, the magnetism is disappeared, and the system becomes pure superconductivity. In Fig.~\ref{fig3}(a),
we also notice that both $\Delta^{s}_{xy}$ and $\Delta_{z}$ versus dopings $x$ display a clear dome-shaped superconductivity, and the values
of $\Delta^s_{xy}$ and $\Delta_{z}$ reach their maximum at the point of the disappearance of magnetic order. The absolute value of
$|\Delta^{d}_{xy}|$ for the two different Fe sublattices is also shown in Fig.~\ref{fig3}(a).

Next, we turn to discussing the behaviors of superfluid density based on the linear response approach. Assuming that, in the
presence of a slowly varying vector potential along the $x$ direction $A_x(r,t)=A(q,\omega)e^{\mathrm{i}\bf{q}\cdot r_i-\mathrm{i}\omega t}$,
all self-consistent mean-field calculations are unchanged in the framework of the linear response theory, only hopping energy terms should be
modified by a Peierls phase factor, $c^{\dag}_{i\sigma}c_{j\sigma}\rightarrow
c^{\dag}_{i\sigma}c_{j\sigma}\exp{\mathrm{i}\frac{e}{\hbar
c}\int^{r_i}_{r_j}\textbf{A}(\textbf{r},t)\cdot \mathrm{d} \textbf{r}}$. Then expanding the factor to the order of $\mathbf{A}^2$, the
perturbed Hamiltonian reads $H^{\prime}=-\sum_iA_x[eJ^P_x(r_i)+\frac{e^2}{2}A_xK_x(r_i)]$ with
\begin{eqnarray}\label{5}
K_x(r_i)&=&-\sum_{\nu\nu^{\prime}\sigma\delta}t_{i,i+\delta}x^2_{i,i+\delta}(c^{\dag}_{i\nu\sigma}c_{i+\delta,\nu^{\prime}\sigma}+H.c.),\\
J^P_x(r_i)&=&-\mathrm{i}\sum_{\nu\nu^{\prime}\sigma\delta}t_{i,i+\delta}x_{i,i+\delta}(c^{\dag}_{i\nu\sigma}c_{i+\delta,\nu^{\prime}\sigma}-H.c.),
\end{eqnarray}
where $\delta=x,x\pm y,x\pm y\pm z$. The total current density $J^{Q}_{x}(r_i,t)=-\frac{\delta H^{\prime}}{\delta
A_x(r_i,t)}$ induced by an external magnetic field is the summation of the diamagnetic part $K_x$ and the paramagnetic part $J^{p}_x$. The calculations
of $K_x$ are restricted to the zeroth order of $A_x(r_i)$ and that of $J^{P}_{x}(r_i)$ is restricted to the first order of $A_x(r_i)$,
$\langle J^{P}_{x}(r_i)\rangle=-\frac{eA_x(r,t)}{N_s}\Pi(\bf{q},\omega)$, where $\Pi(\textbf{q},\omega)$ is obtained from the analytic
continuation of the current-current correlation $\Pi(\textbf{q},\mathrm{i}\omega)=\int^{\beta}_0 d\tau
e^{\mathrm{i}\omega\tau}\Pi({\bf{q}},\tau)$ in the Matsubara formalism. Here $\Pi(\textbf{q},\tau)=-\langle T_{\tau}
J^P_x(\textbf{q},\tau)J^P_x(-\textbf{q},0)\rangle_0$, $J^P_x(\textbf{q},\tau)\,=e^{\tau H_0}J^P_x(\textbf{q}) e^{-\tau H_0}$,
$J^P_x(\textbf{q})=\sum_{i}e^{-\mathrm{i}\textbf{q}\cdot\textbf{r}_i}J^P_x(r_i)$, and $T_{\tau}$ is the imaginary time ordering operator. In
the quasi-particle basis, the paramagnetic current can be expressed as the summation of components $J^P_{x}({\bf{q}})=\sum_{m_1m_2}J^P_{m_1m_2}$
with $J^P_{m_1m_2}=\gamma^{\dag \bf{k}}_{m1}\gamma^{ \bf{k+q}}_{m2}\Gamma^{\bf{k,k+q}}_{m_1m_2}$. After some tedious but straightforward
algebraic derivations, the concrete expression of $\Gamma$ is derived. Using the equation of motion of Green's function, we
obtain~\cite{huang2}
\begin{eqnarray}\label{corr}
\Pi(\textbf{q},\mathrm{i}\omega)
            \!\!=\!\!\!\!\!\sum_{{\bf{k}}m_1m_2\sigma} \!\! \frac{
            \Gamma_{m_1m_2}^{\bf{k},\bf{k+q}}\Gamma_{m_2m_1}^{\bf{k+q},\bf{k}}[f(E_{{\bf{k}},m_1})-f(E_{{\bf{k+q}},m_2})]
            }{\mathrm{i}\omega+(E_{{\bf{k}},m_1}-E_{{\bf{k+q}},m_2})},
\end{eqnarray}
where $f$ is the Fermi-Dirac distribution function. Thus, the superfluid density weight measured by the ratio of the superfluid density to the
mass is proportional to $\Pi(\bf{q},\omega)$ in the limit of zero frequency $\omega$ and momentum $\bf{q}$~\cite{huang2,tdas,djs,fla2} and is
expressed as
\begin{eqnarray}\label{13}
&&\frac{\rho_s}{m^{\ast}}=-\langle J^{Q}_{x}(r_i,t) \rangle/e^2A_x(r_i)\nonumber\\
&=&\frac{1}{N_s}\Pi(q_x=0,q_y\rightarrow 0,q_z\rightarrow 0,\omega=0)-\langle K_x \rangle_0.
\end{eqnarray}

Fig.~\ref{fig3}(b) shows the superfluid density $\rho_s$ as a function of doping concentration $x$ across the whole phase diagram in
Fig.~\ref{fig3}(a) at zero temperature. At underdoped concentration $x$ around the parent compound system, the superfluid density $\rho_s$ is
zero as expected from our intuitive knowledge that the system does not have superconducting order. As the doping concentration $x$ increases,
the superconductivity emerges, accompanied by the appearance of a finite value of $\rho_s$. If the doping concentration $x$ is further
increased, $\rho_s$ changes to decrease its value slowly until $x=0.22$, and then $\rho_s$ further goes upward with a steep slope, displaying a
sharp peak at $x=0.3$ with the value of superfluid density being $8$ times larger than that at $x=0.22$. Reaching the maximal value of
superfluid density $\rho_s$ at $x=0.3$ corresponds to the point of the disappearance of magnetic order in Fig.~\ref{fig3}(a), denoted
by the red dot in the curve of $\Delta^s_{xy}$. Eventually, the superfluid density $\rho_s$ decreases sharply and then tends to saturates to a
finite value upon further increasing the doping concentration $x$.

In addition, a fundamental property of the superconducting state is the London penetration depth $\lambda$, parametrizing the ability of a
superconductor to screen an applied magnetic field, which not only can be evaluated straightforwardly from the superfluid density $\rho_s$ but
also can be measured in experiments~\cite{yla}. In general, $\rho_s$ is described as the phase rigidity of a superconductor, and it may vanish
before the superconducting energy gap diminished as increasing temperature. In Fig.~\ref{fig3}(c), we plot the square of London penetration
depth $\lambda^2$ as a function of doping concentration $x$. It is important to point out that the value of London penetration depth
$\lambda^2$ displays a sharp peak at $x=0.22$ which corresponds to the minimal value of $\rho_s$ and corresponds to the red square in the curve
of $\Delta^s_{xy}$ in Fig.~\ref{fig3}(a), and then it decreases sharply. Eventually, $\lambda^2$ becomes rather flat in the pure
superconducting region. Compared with the experimental results~\cite{Hashimoto}, where the magnetic phase boundary corresponds to the sharp peak of $\lambda^2$, our numerical results show that the sharp peak penetration depth appears before the vanishing of magnetic order. Such anomalous
peak in London penetration depth has never been observed experimentally in other iron-based superconductors, and it leads to a conjecture of
the presence of QCP in BaFe$_2$(As$_{1-x}$P$_x$)$_2$.

\begin{figure}
\centering
\includegraphics[width=8.0cm]{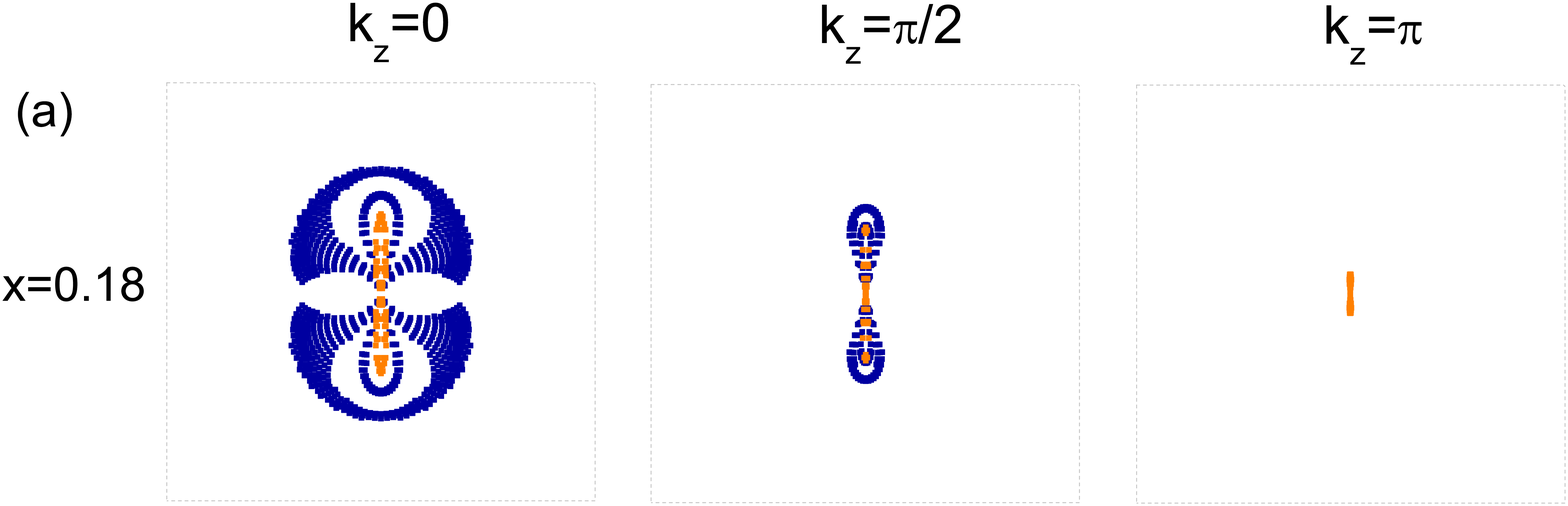}
\includegraphics[width=8.0cm]{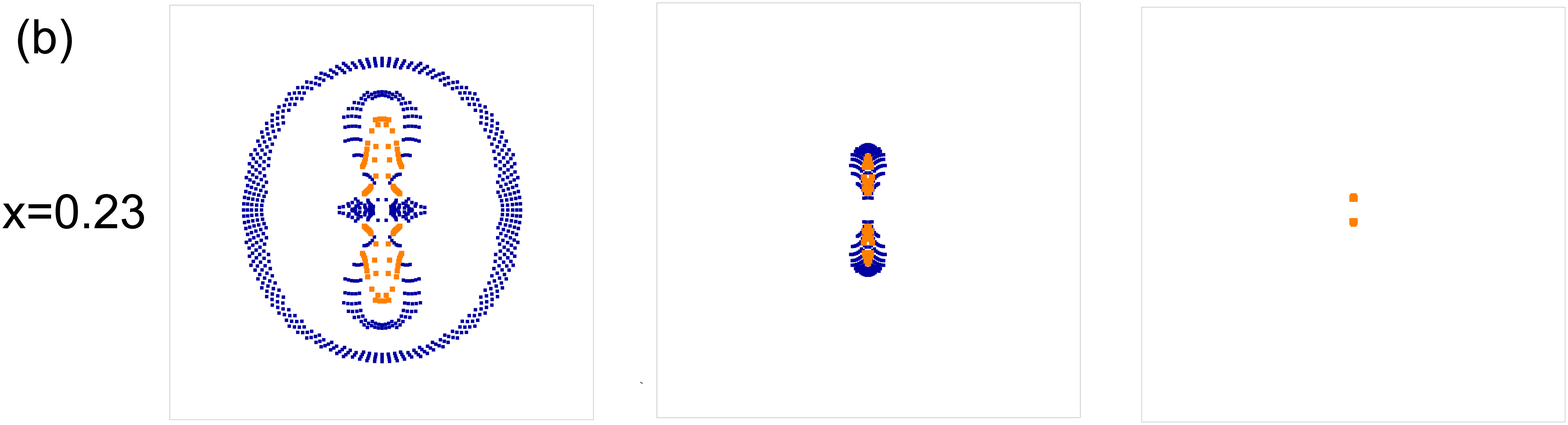}
\includegraphics[width=8.0cm]{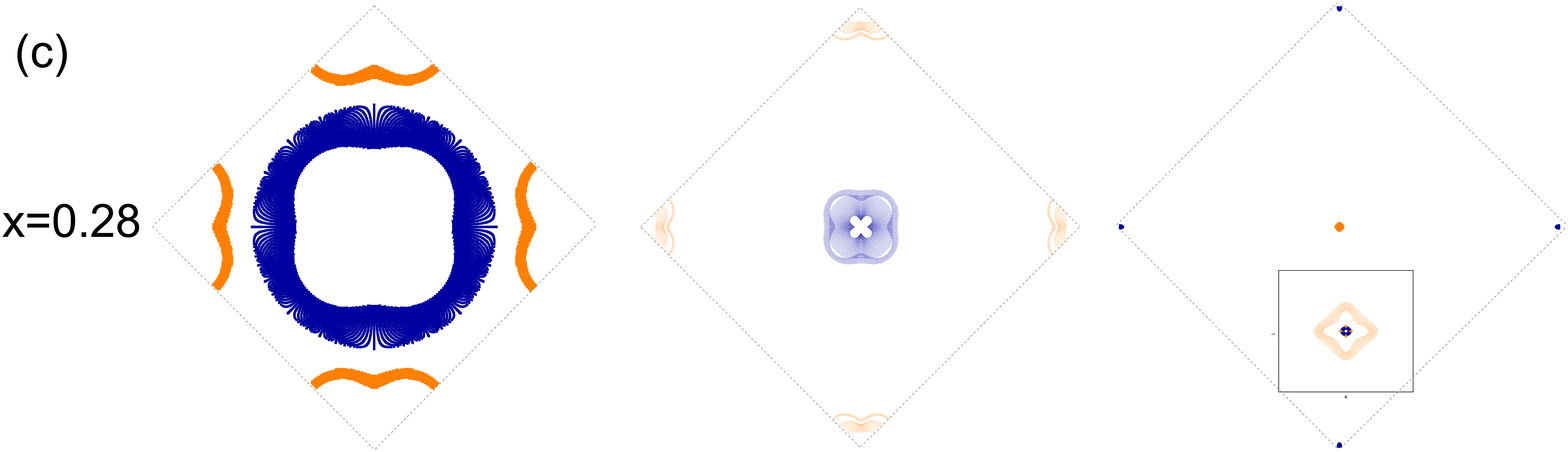}
\caption{(Color online) Superconducting gap structure of the band space for different doping concentrations $x=0.18,0.23,0.28$ near the Fermi surface. From left to right the plots correspond to  $k_z=0,\pi/2,\pi$, respectively. The inset of panel (c) is an enlargement of the $z$ point. The gray lines are the corresponding first Brillouin zone.}\label{fig4b}
\end{figure}

The superconducting gap structure in the band representation $\tilde{\Delta}_k$ is derived from the $6\times 6$ matrix Hamiltonian $H_{\Delta,k}$ of
$\Delta_k$ (short hand for $\Delta_{R\nu k}$) in the orbital space as $[\tilde{H}_{\Delta,k}]=[W]^{\dag}[H_{\Delta,k}][W]^{*}$ when magnetic order is absent, with
$[W]$ being the transformation matrix diagonalizing the $6\times 6$ tight-binding Hamiltonian $H_{t,k}$, and the corresponding $\tilde{\Delta}_k$
are the diagonal elements of $[\tilde{H}_{\Delta,k}]$. However, for finite magnetic order the corresponding $[W]$ is a $12\times 12$ matrix diagonalizing the Hamiltonian $H_{t,k}+H_{int,k}$ including the interaction part. Figure~\ref{fig4b} displays the behavior of $\tilde{\Delta}_k$ near the Fermi surface, where the navy points correspond to positive signs of $\tilde{\Delta}_k$ and the orange points are for the minus signs.
Figure~\ref{fig4b} shows that the gap structure in $x=0.23$ has a finite value near the Fermi surface at all $k_z$, which is quite different from that in the $x=0.18$ case where the node points exist. For a given doping level, a larger $k_z$ corresponds to a smaller magnitude of superconducting gap.
It is important to point out that along the line of $k_z$, the superconducting gap will change signs from $0.5\pi$ to $\pi$ when we do not consider the effect of magnetic order, which can be seen clearly in Fig.~\ref{fig4b}(c), where we set the magnetic order to zero.
Therefore, we expect that in the region with the sudden drop in penetration depth the corresponding $\tilde{\Delta}_k$ will change its structure.

\section{DOS and Fermi surface topologies}\label{dos}

\begin{figure}
\centering
\includegraphics[width=8.0cm]{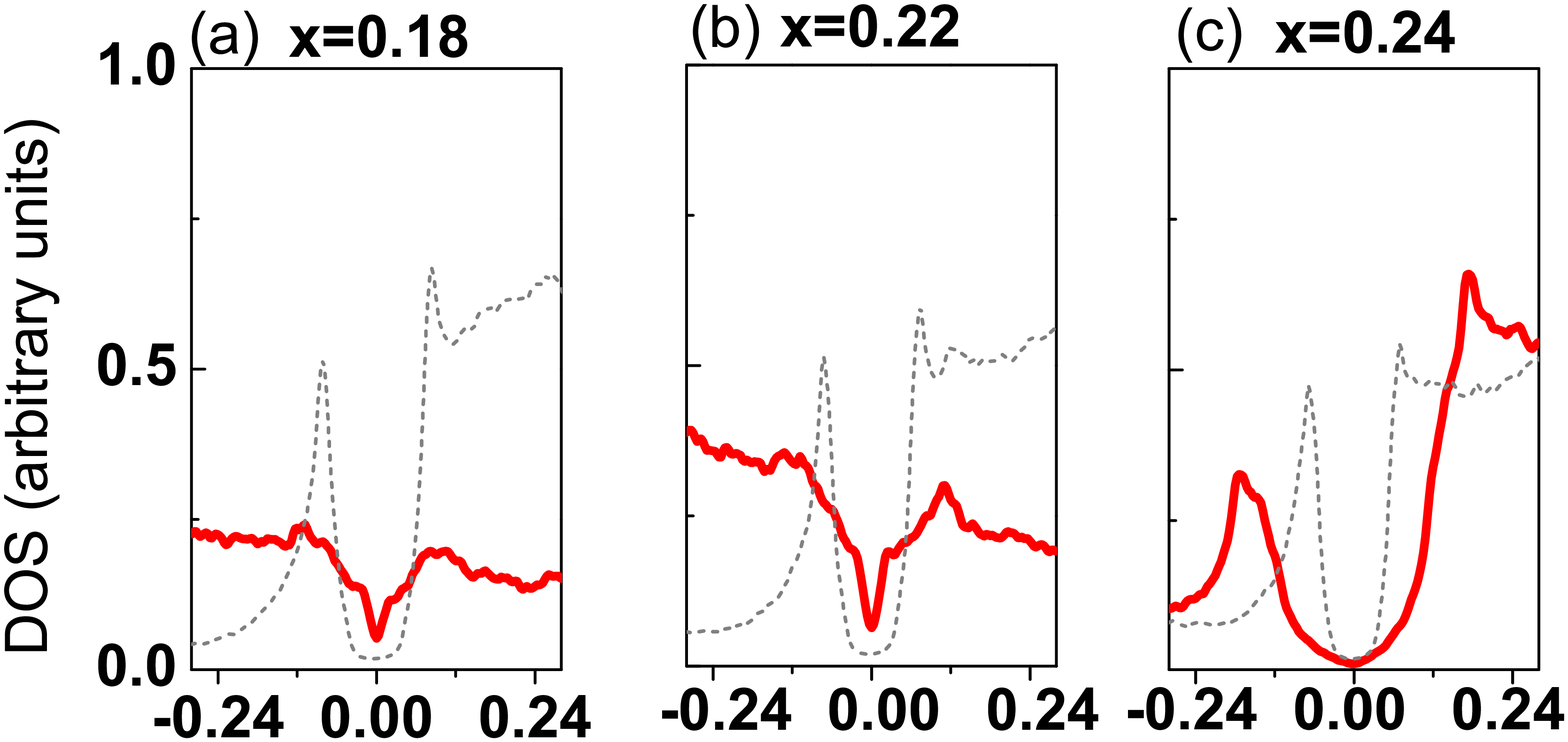}
\includegraphics[width=8.0cm]{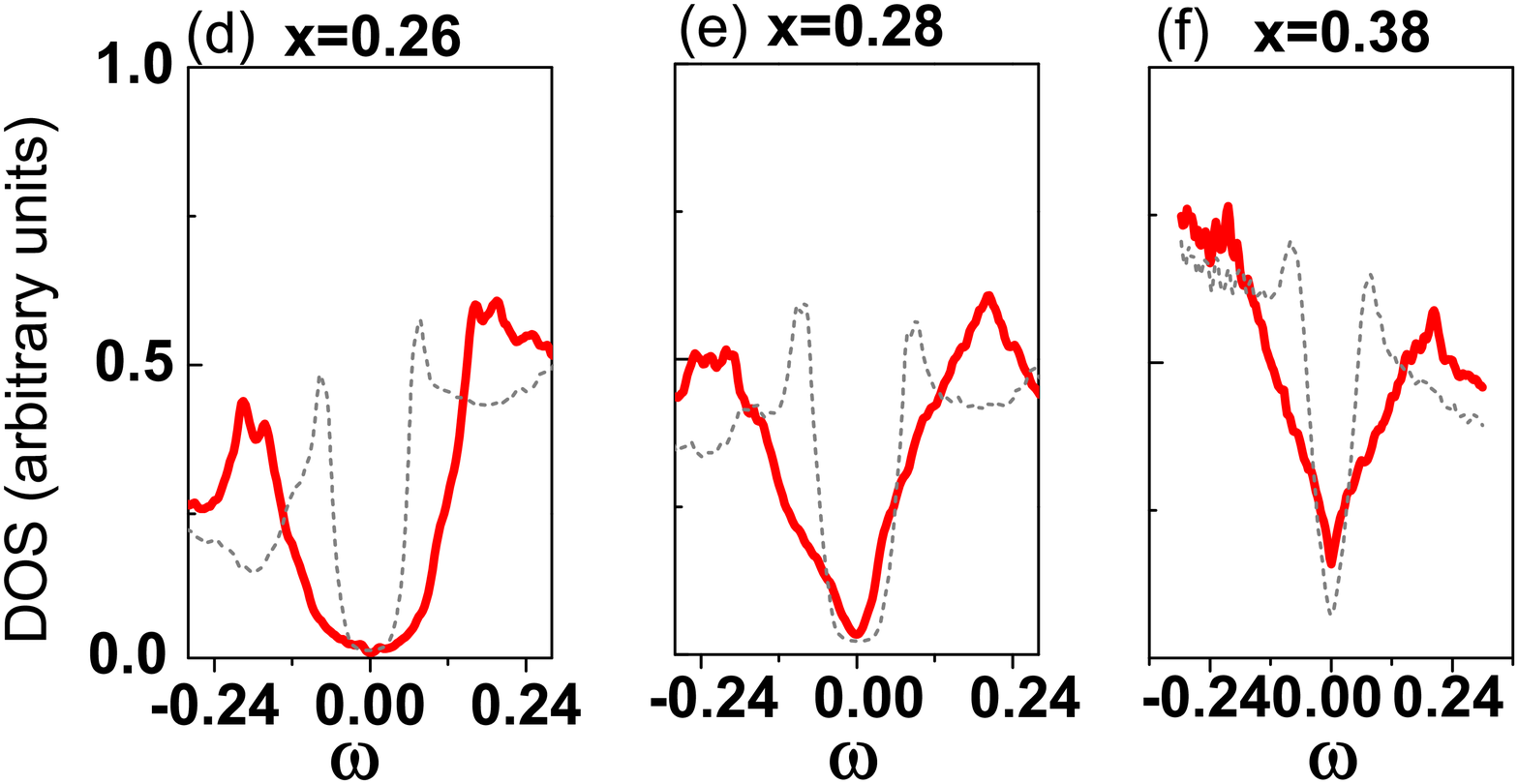}
\caption{(Color online) (a)-(f) For $V_z=1.6$, DOS as a function of energy at various doping concentrations
$x=0.18,0.22,0.24,0.26,0.28,0.38$ are shown by the red lines, where the gray dashed lines represent the corresponding DOS for
$V_z=0$.}\label{fig4}
\end{figure}

To clarify the nature of the emergence of an anomalous sharp peak in the penetration depth, we calculate the DOS for various doping
concentrations $x$ at zero temperature, as shown in Fig.~\ref{fig4}. When the doping concentration is located in the region of $x\leq0.22$, the
calculated DOS displays a ``V"-shaped structure with finite value at the Fermi level, implying the presence of nodal points in the
superconducting energy gap. As the doping concentration $x$ increases, the ``V"-shaped DOS changes into a ``U"-shaped structure at
$x=0.24,0.26$ with a diminished DOS at the Fermi level; the plot for $x=0.23$ is similar to that for $x=0.24$, which we do not show here. When the doping concentration is increased beyond $0.28$, the tip at zero energy
reappears [see Fig.~\ref{fig4}(e)]. Fig.~\ref{fig4}(f) shows a narrow ``V"-shaped DOS feature in a pure superconducting region,
suggesting the system is a nodal superconductor, which is in agreement with the previous ARPES measurement~\cite{YZhang}.
Therefore, comparing Fig.~\ref{fig4} with Fig.~\ref{fig3}(c) we find that the phase transition of the changing pairing order parameter from nodeless to a nodal
structure is responsible for the appearance of the sharp peak in the experimental measurement of the London penetration depth.

To further understand the nature of the emergence of the anomalous sharp peak in the London penetration depth, we also plot the DOS for
$\Delta_z=0$, a two-dimensional limited case, shown by the gray dashed lines in Fig.~\ref{fig4}, where the interlayer interaction $V_{z}=0$ is set to
zero and the other interaction parameters remain the same as in Fig.~\ref{fig2}. For this $3D$ interaction and two-dimensional superconducting
order case, the system displays a ``U"-shaped DOS for all doping concentrations from $x=0.18$ to $x=0.28$.

\begin{figure}
\centering
\includegraphics[width=9cm]{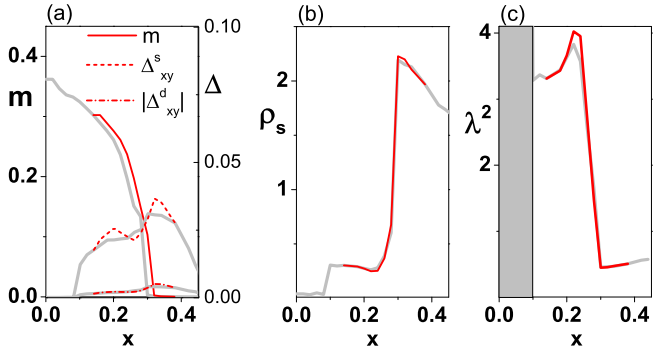}
\caption{(Color online) In the doping region of [0.14,0.36], the red curves are doping-dependent physical quantities for the $\Delta_{z}=0$ case.
(a) Magnetic order $m$ and superconducting orders $\Delta^{s,d}_{xy}$, (b) superfluid density $\rho_s$, and (c) the square of the penetration depth as
a function of doping. The gray lines are the corresponding  curves of $V_{z}=1.6$ in Fig.\ref{fig3}.
}\label{fig9}
\end{figure}

In the case where $\Delta_z$ is absent, the nodal-to-nodeless transition no longer exists since all the DOS are ``U"-shaped features.
Although the resulting $\rho_s$ and $\lambda^2$ still have a sharp peak as that in $V_z=1.6$ case, which can be seen clearly in Fig.~\ref{fig9}(b) and
Fig.~\ref{fig9}(c), a remarkable dip in the phase diagram of $\Delta^s_{xy}$ versus doping appears, which is ascribed to the presence of three-dimensional interaction.
Figure~\ref{fig9}(a) shows that $\Delta^s_{xy}$ drops to a minimum value suddenly at $x=0.24$, destroying
the
dome-shaped superconductivity and leading to an unphysical anomalous penetration depth. A minimum pairing order corresponding to a maximal
penetration depth is a reasonable result when there is no other phase transition.
Furthermore, for two-dimensional dome-shaped iron-based superconductivity~\cite{huang2}, the penetration depth does not show the sharp peak.
Therefore, the experimental observation of a sharp peak in penetration depth having dome-shaped superconductivity stems from three-dimensional
electron interactions accompanied by a transition from nodal to nodeless pairing.

\begin{figure}
\centering
\includegraphics[width=2.4cm]{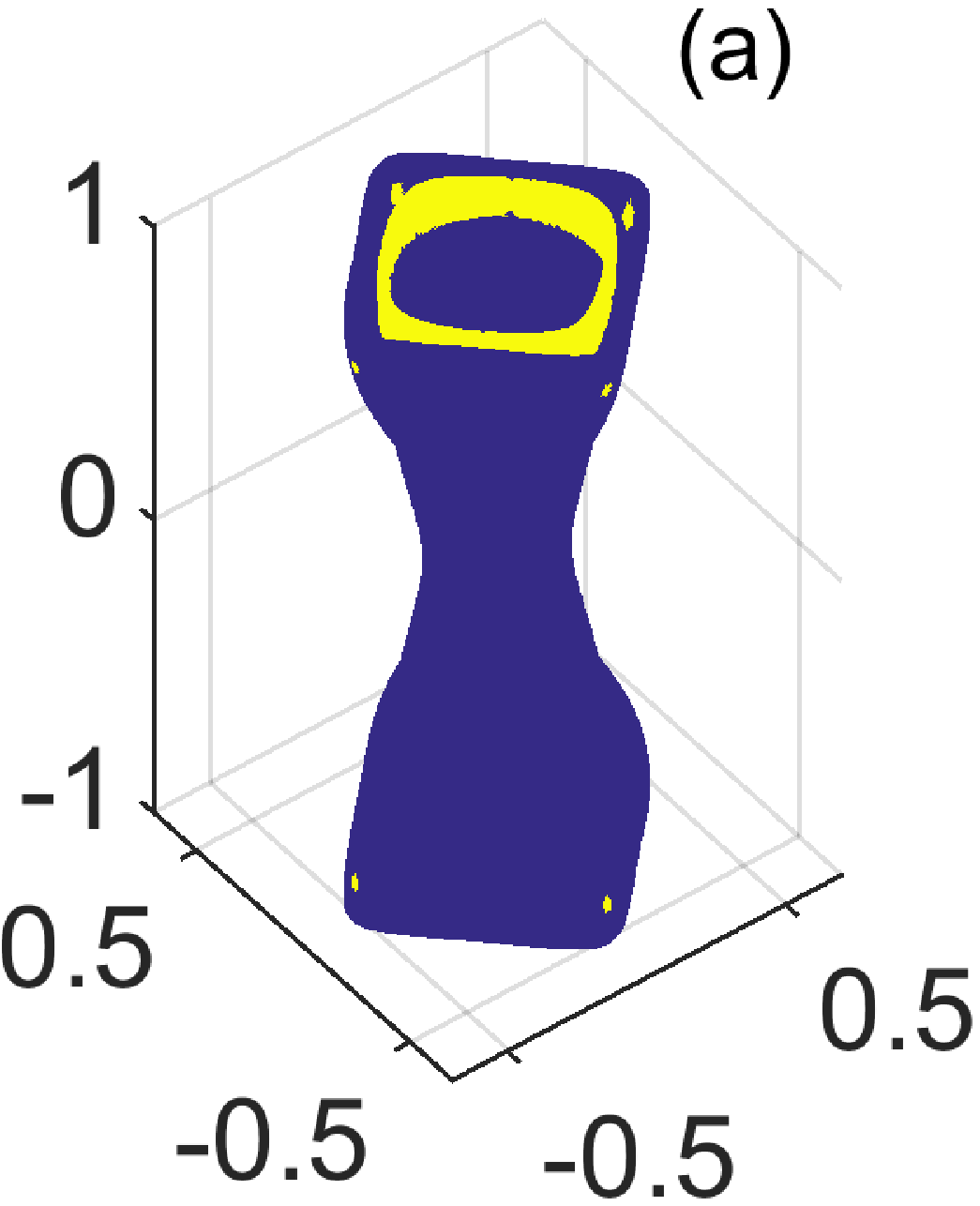}
\includegraphics[width=2.7cm]{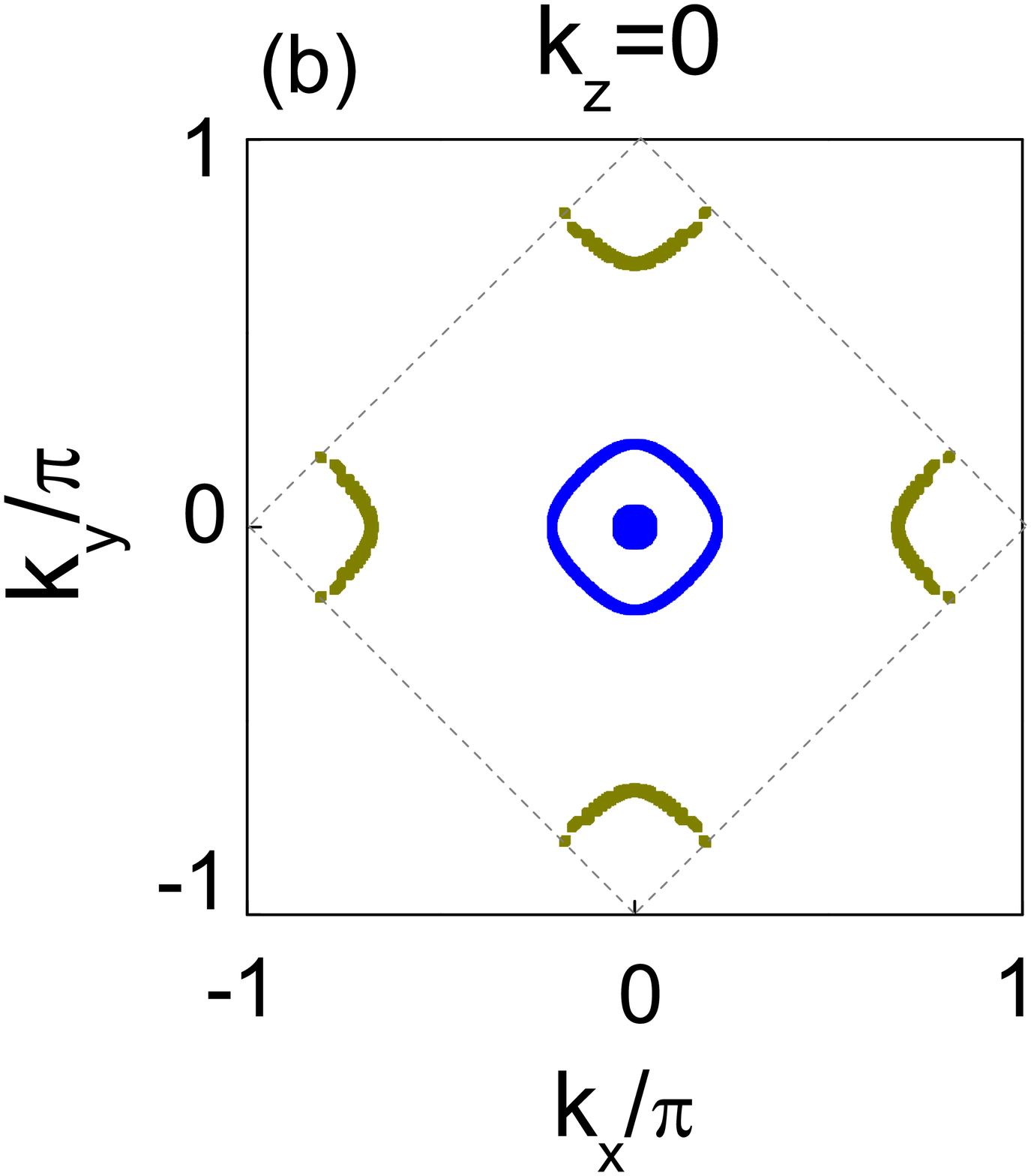}
\includegraphics[width=2.7cm]{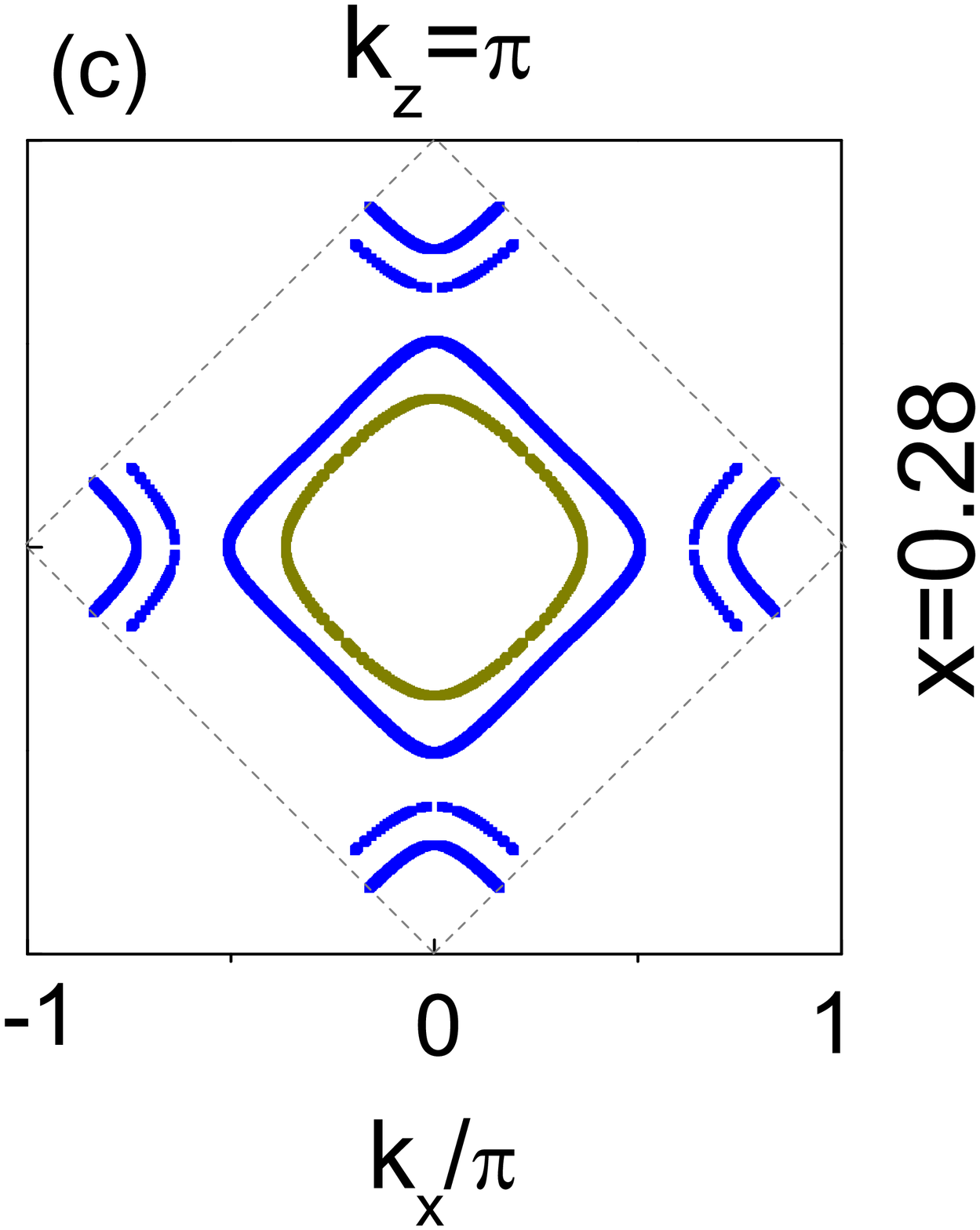}
\caption{(Color online) (a) Three-dimensional plot of the two cylindrical shells around the $\Gamma$ point for dopant concentration of $x=0.28$.
(b) and (c) are the contour plots of the Fermi surface at $k_z=0$ and $k_z=\pi$, respectively. Blue points denote the plus signs of the
corresponding superconducting pairing, and dark yellow points are for the minus signs.}\label{fig5}
\end{figure}

In addition to analyzing the numerical data for superconducting pairing order parameters, we find a nodal circle in the inner hole
pocket  around the $\Gamma$ point in the vicinity of $k_z=0.86\pi$ for $x=0.28$, and four nodal points in the outer hole pocket, which is
shown in Fig.~\ref{fig5}(a) with the boundaries of the two colors denoting the nodal points on the $3D$ Fermi surface topologies. For
$k_z>0.86\pi$, the superconducting pairing in the inner hole pocket changes sign, while the points on the outer pocket remain the same
color as that of small $k_z$, which is quite different from $x=0.18$ shown in Fig.~\ref{fig1}(b).
In order to unambiguously display the inner shape of Fermi surface topologies, Figs.~\ref{fig5}(b) and (c) depict the contour plot of Fermi
surface for $k_z=0 $ and for $k_z=\pi$ at $x=0.28$, respectively. For $k_z=\pi$ the superconducting gap has different signs on the outer and
inner hole pockets, which is different from $x=0.18$ but consistent with the inset of Fig.\ref{fig4b}. It is worth pointing out that the small
inner circle of the hole pocket around the $\Gamma$ point in the Brillouin zone is easily immersed if the doping concentration is
increased further. The larger $k_z$ has a larger Fermi surface circle; however the magnitude of the corresponding superconducting pairing is small as
shown in Fig.~\ref{fig4b}. Those calculations further solidify the nature of nodeless-to-nodal transition in doped
BaFe$_2$(As$_{1-x}$P$_x$)$_2$, leading to the appearance of an anomalous sharp peak in the London penetration depth.

\section{summary}\label{final}

In this paper, we constructed a three-dimensional tight-binding lattice model based on the facts from the penetration depth and the ARPES
experimental measurements. Taking the interlayer Coulomb interactions into account, the superconducting phase diagram and an anomalous sharp
peak in the London penetration depth were evaluated, and are entirely in good agreement with experimental observations. By verifying the DOS and
the pairing order parameters as well as the Fermi surface topologies at various doping concentrations, we find that the QCP originates from the
nature of three-dimensional interactions, leading to a phase transition from a nodeless to a nodal pairing symmetry. This finding provides
significant insight into the understanding of the nature of the QCP that emerged in the London penetration depth experiment in the isovalent doped
superconductor BaFe$_2$(As$_{1-x}$P$_x$)$_2$.

\section{acknowledgments}
This work was supported by the National Key Research and Development Programs of China (Grant Nos. 2017YFA0304204 and 2016YFA0300504), Nat Basic Research Program of China (Grant No. 2914CB921203), the National Natural Science Foundation of China (Grant Nos. 11625416, 11474064, 11674278, 11927807 and 11774218),Shanghai Municipal Government under the Grant No. 19XD1400700, the Natural Science Foundation of Shanghai of China (Grant Nos. 18JC1420402 and 19ZR1402600), and the Natural Science Foundation from Jiangsu Province of China (Grant No. BK20160094). W. L. also acknowledges the start-up funding from Fudan University.

\end{document}